\documentclass[aps,prl,a4paper,reprint,showpacs,amsmath,amssymb,superscriptaddress,floatfix]{revtex4-1}

\usepackage{amssymb}
\usepackage{latexsym}
\usepackage[dvips]{graphicx}
\usepackage{epsfig}

\usepackage{dcolumn}
\usepackage{amsmath}
\usepackage{amsfonts}
\usepackage{bm}
\usepackage{color}

\newcommand{\be}{\begin{equation}}
\newcommand{\ee}{\end{equation}}
\newcommand{\bea}{\begin{eqnarray}}
\newcommand{\eea}{\end{eqnarray}}

\newcommand{\p}{\partial}

\newcommand{\hs}{\hat{\sigma}}

\newcommand{\rcd}{{\cal D}}
\newcommand{\ri}{\mbox{i}}
\newcommand{\re}{\mbox{e}}

\begin{document}
\title{Quantum Phase Transition and Protected Ideal Transport in a Kondo Chain}
\author{A. M. Tsvelik}
\affiliation{Brookhaven National Laboratory, Upton, NY 11973-5000, USA}

\author{O.M. Yevtushenko}
\affiliation{Ludwig Maximilians University, Arnold Sommerfeld Center and Center for Nano-Science, Munich, DE-80333, Germany}

 \date{\today }

\begin{abstract}
We study the low energy physics of a Kondo chain where electrons from a one-dimensional band interact with
magnetic moments via an anisotropic exchange interaction. It is demonstrated that the anisotropy
gives rise to two different phases  which are separated by a quantum phase transition. In the phase with
easy plane anisotropy, Z$_2$ symmetry between sectors with different helicity of the electrons is broken.
As a result, localization effects are suppressed and the dc transport acquires (partial) symmetry
protection.
This effect is similar to the protection of the edge transport in time-reversal
invariant topological insulators. The phase with easy axis anisotropy corresponds to the Tomonaga-Luttinger
liquid with a pronounced spin-charge separation. The slow charge density wave modes have no
protection against localizatioin.
\end{abstract}


\maketitle
{\it Introduction}. One-dimensional systems present an ideal platform for formation of charge density waves (CDW)
\cite{Giamarchi};  the transport in clean systems is almost  ideal \cite{rosch}. However, for realistic interactions and
at low temperatures,
even a weak disorder
pins the CDW suppressing the charge transport \cite{GiaSchulz}. The ideal transport can be protected by symmetries:
a well-known example is  the edge transport in two-dimensional time-reversal invariant topological
insulators (TIs)\cite{HasanKane,QiZhang,TI-Shen,MolFranz}.
The topologically non-trivial state of the bulk and time-reversal symmetry lead to a lock-in relation between
the chirality and the spin of edge modes making them helical \cite{WuBernevigZhang}.
As a result, the electron backscattering
must be accompanied by a spin-flip;  hence the edge transport becomes immune to effects of potential disorder.
Other processes which can  suppress the ideal transport  include scattering by magnetic impurities \cite{FurusakiMatveev}
or inelastic processes due to interactions \cite{FvOpp-Inelast,CheiGlaz,GoldstGlazman-Short,GoldstGlazman-Long,Mirlin-HLL}.
All of them become ineffective at low temperatures. The presence of  (almost) ballistic edge transport has been
confirmed in state-of-the-art experiments
\cite{Molenkamp-2007,Molenkamp-2009,EdgeTransport-Exp0,InAs-GaSb-Meas}.  Hence it is accepted that the
ballistic transport is protected by time-reversal symmetry and this protection is removed when this symmetry
is broken \cite{AAY,Yevt-Helical}.

Helical boundary modes can exist in noninteracting systems due to topological nontriviality of the bulk
\cite{TITS-book}. {\it In this Letter}, we show that  helical modes
may emerge in interacting systems as a result of spontaneous symmetry breaking.
As an illustration, we study a model of Kondo
chain \cite{ZachEmKiv,KL-Honner-Gulacsi,KL-Rev1,KL-CoulGas,review-gulacsi} consisting of band
one-dimensional electrons interacting with local spins; the Hamiltonian of this system is:
\be
\label{KCh}
   \hat{H} = -  t \sum_{n} \hat{c}^\dagger_{n+1} \hat{c}_n +
                        \sum_m J_a \, \hat{c}^\dagger_m \, \hs^a \hat{S}^a(m) \, \hat{c}_m  + H.c.
\ee
Here $ \, \hat{c}^{\rm T}_n \equiv (\hat{c}_{\uparrow}(n),  \hat{c}_{\downarrow}(n)) \, $ are electron operators at lattice
site $ \, n $; $ \, \hs^a \, $ are Pauli matrices ($ a = x, y, z$); $ \, \hat{S}^a(m)$ are components of the spin-s operator
located on lattice site $ \, m $; $ \, t \, $ denotes the overlap integral.
It is assumed
that sites $\{m\}$ constitute some (not necessarily regular) subset of sites $\{n\}$. We concentrate on the regime of
sufficiently high density of spins where the Kondo effect is suppressed and the physics  is  determined  mostly by the
Ruderman-Kittel-Kasuya-Yosida (RKKY) interaction \cite{NoKondo}. The band is far from half filling,
the spins are quantum and the coupling constants are much smaller than the bandwidth, $ \, s J_{a} \ll t $.
We will consider the coupling which is isotropic in the $ \, XY $-plane: $ \, J_x = J_y \equiv J_{\perp} $.

{\it Brief summary of the results}:
The low energy (LE) behavior of model (\ref{KCh}) includes
two distinct regimes corresponding to the easy axis (EA), $J_z > J_{\perp}$, and the easy
plane (EP), $J_z < J_{\perp}$, anisotropy. In the first case, all quasiparticle (fermionic) excitations
are gapped. The transport is carried by  gapless collective modes, the charge and the spin density waves.
The CDW couples to a potential disorder which is able to pin it and to block the charge transport.
The SU(2) symmetric point, $ J_z = J_{\perp}$, is the point of quantum
phase transition into a phase with spontaneously broken helicity.  In the EP phase at $T=0$,
quasiparticles with a given helicity acquire a gap and the other helical branch remains gapless.
The charge transport is carried by the gapless helical electrons and by the slow collective
excitations (spin-fermion waves). If the spin U(1) symmetry is respected the long range
helical ordering makes single-particle backscattering of the gapless modes
impossible as in the noninteracting TIs. This leads to suppression of localization
effects: the localization radius becomes parametrically large and the dc transport acquires
a (partial) symmetry protection in finite but long samples.


{\it Continuum limit}: To describe the LE physics we develop a continuum limit theory.
This requires to single out smooth modes. We linearize the spectrum of electrons and
expand operators $  \hat{c} $ in smooth
chiral modes:
\be
  \hat{c}_{ \uparrow \downarrow}(n) =
            \re^{-\ri k_F \xi_0n} \hat{R}_{ \uparrow \downarrow}(x) + \re^{\ri k_F \xi_0n}\hat{L}_{ \uparrow \downarrow}(x), \
   x = n \xi_0 \, ;
\label{ferm}
\ee
were $\xi_0$ is the lattice constant. The Lagrangian  density of the band electrons becomes
\be
\label{BandLagr}
  {\cal L}_{\rm e} = \Psi^\dagger \Big[ (\hat{I} \otimes \hat{I} )\p_{\tau} - \ri ( \hat{I} \otimes \hat{\tau}^z) v_F \p_x \Big]\Psi .
\ee
Here $ \tau $ is the imaginary time;
the first space in the tensor product is the spin one, the Pauli matrices $ \hat{\tau}^a$ act in the chiral space;
$ \hat{I} = {\rm diag}(1,1) $; $ \, v_F = 2 t  \xi_0 \sin( k_F \xi_0 ) $ is the Fermi velocity ($ \, k_F \, $ is the Fermi
momentum); $ \, \Psi^{\rm T} = \left( R^{\rm T}, L^{\rm T} \right) \, $ is the 4-component fermionic spinor field.

Contrary to Ref.\cite{ZachEmKiv}, where the effects of forward scattering at $ J_z \sim t $  (i.e., of the Kondo
physics)  were considered, we suggest that the LE physics in the dense limit with $ J_a \ll t $ (dominated by the
RKKY interaction) is governed by backscattering of the fermionic modes.  It is described by
\be
\label{CouplLagr}
   {\cal L}_{\rm bs} = \rho_s \sum_{a=x,y,z} J_a \sum_{m} \re^{2 \ri k_F \xi_0 m} R^\dagger S^a(m) \hs^a L + H.c.
\ee
$ \rho_s $ denotes the dimensionless spin density. $  {\cal L}_{\rm bs} $ is expected to lead to opening of the
spectral gaps thus reducing  the energy of the electrons. As will be clear from the subsequent discussion, the
resulting physics is quite different from that of Ref.\cite{ZachEmKiv}.

\begin{figure}[t]
   \includegraphics[width=0.3 \textwidth]{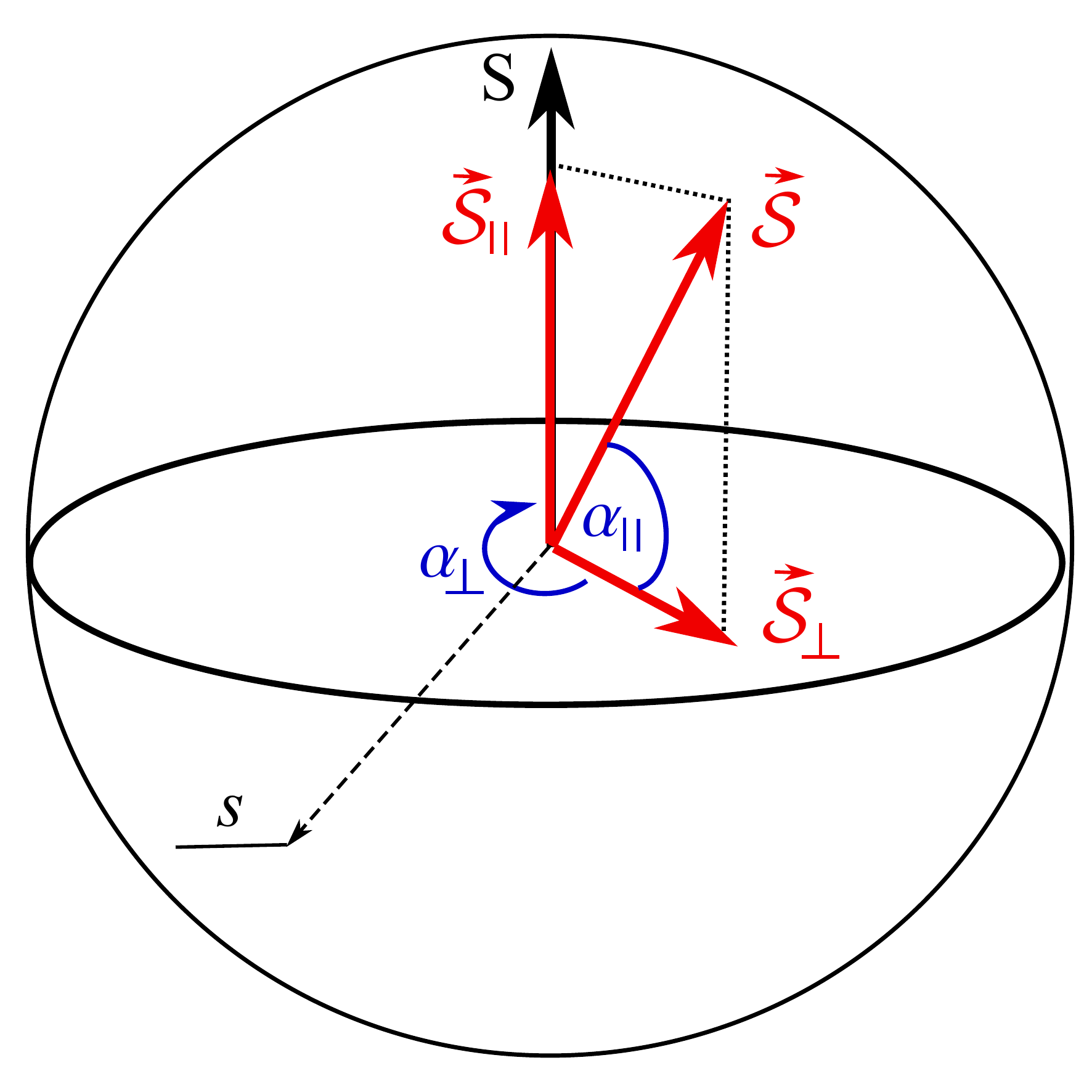}
   \caption{\label{Geom}
        Transformation from the frame of the vector $ {\bf S} $ to that of $ \roarrow{\cal S} $.
        Angles $  \alpha_{\parallel,\perp} $ define the modulus of the transverse component
        $ \roarrow{\cal S}_\perp $ and its rotation around the longitudinal component $ \roarrow {\cal S}_\parallel $,
        respectively.
                 }
\end{figure}

We can eliminate the oscillatory factors in (\ref{CouplLagr}) by absorbing them into the spin
configurations which amounts to separation of  fast and slow spin variables  \cite{HelicConfig}.
The standard parametrization of the spin by azimuthal and polar
angles, $ {\bf S} = s \{ \sin(\theta)\cos(\psi), \sin(\theta)\sin(\psi), \cos(\theta) \} $, with the
integration measure $ {\cal D} \{ \Omega_S \} = \sin(\theta) {\cal D} \{ \theta \} {\cal D} \{ \psi \} $
\cite{Lwz} is not convenient for our purposes. Therefore,
we change to the rotating orthonormal basis $ {\bf e}_{1,2,3} $
with $ {\bf e}_{3} = {\bf S}/s $. We define a ``longitudinal'', $ \roarrow {\cal S}_\parallel $, and the ``transverse'',
$ \roarrow{\cal S}_\perp $, components of the new spin vector $ \roarrow{\cal S} = \roarrow{\cal S}_\perp +
\roarrow {\cal S}_\parallel $ (Fig.\ref{Geom}):
\be
\label{NewSpin}
    \frac{\roarrow {\cal S}_\parallel}{s} \equiv  {\bf e}_3 \sin\alpha_\parallel ; \,
    \frac{\roarrow{\cal S}_\perp}{s}  \equiv [  {\bf e}_1 \cos \alpha_\perp + {\bf e}_2 \sin \alpha_\perp  ] \cos \alpha_\parallel ;
\ee
$ \alpha_\perp = 2 k_F \xi_0 m + \alpha (x) $. The orthonormality can be resolved by choosing
\bea
\label{Evec-1}
   {\bf e}_1 & = & \{ -\cos(\theta) \cos(\psi), -\cos(\theta) \sin(\psi), \sin(\theta) \} , \\
\label{Evec-2}
   {\bf e}_2 & = & \{ \sin(\psi), -\cos(\psi), 0\} .
\eea
The integration measure for $  \alpha,\alpha_{\parallel} $ will be
$ {\cal D} \{ \Omega_\alpha \} = \cos(\alpha_\parallel) {\cal D} \{ \alpha_\parallel \} {\cal D} \{ \alpha \}  $,
the total measure reads $ {\cal D} \{ \Omega \} =  {\cal D} \{ \Omega_\alpha \} {\cal D} \{ \Omega_S \} $.
This does not result in overcounting  the degrees of freedom  since
we will find a scale separation with two fast (massive $ \alpha_\parallel, \theta $) and two slow
(massless $ \alpha, \psi $) angles \cite{ScaleSep}.
Verification of the scale separation and stability of the chosen spin configuration will confirm self-consistency
of our approach.

Inserting the new parametrization in Eq.(\ref{CouplLagr}) and keeping only the non-oscillatory terms, we find
LE Lagrangian $ {\cal L}_{\rm eff} = {\cal L}_{\rm e} + {\cal L}_{\rm bs}^{\rm (sl)} +  {\cal L}_{\rm WZ} $ where
\bea
\label{CouplLagrSlowParam}
   {\cal L}_{\rm bs}^{\rm (sl)} & = & \frac{\tilde{s} \rho_s}{2}
             R^\dagger \Bigl\{
                 J_\perp \left[ \re^{\ri \psi} \sin^2 \! \left( \frac{\theta}{2} \right)\hs^-  \!\! -
                                       \re^{-\ri \psi} \cos^2 \! \left( \frac{\theta}{2} \right)\hs^+
                               \right] \cr
                                 & & + J_z \sin(\theta)\hs^z \Bigr\} L \re^{-\ri \alpha} + H.c. ; \quad
   \tilde{s} \equiv s \cos(\alpha_\parallel) ;
\eea
$ {\cal L}_{\rm WZ} $ is
the topological Wess-Zumino term
\cite{ATsBook,Suppl1}:
\bea
   {\cal L}_{\rm WZ} & = & \ri s \rho_s \xi_0^{-1} \sin(\alpha_\parallel) [  \p_\tau \alpha + \cos(\theta) \p_\tau\psi  ] .
   \label{WZ}
\eea
The fermionic gaps become maximal at $ \alpha_{\parallel} = 0 $ and $ \theta = 0, \pi/2, \pi $. Thus,
we expect three extrema of the action whose stability depends on the ratio $ J_\perp / J_z $.

{\it EA anisotropy, $ J_z > J_\perp $}:
The term $ O(J_z) $ dominates and opens the gap
in all fermionic modes. This can be shown straightforwardly after removing the angles $\alpha,\psi$ from
the backscattering term (\ref{CouplLagrSlowParam})
by using the Abelian bosonization \cite{Bosonisation,Funcb}:
we bosonize the fermions and shift bosonic phases:
\be
\label{Shift}
   \tilde{\Phi}_c = \Phi_c - \alpha/2, \
   \tilde{\Theta}_s =  \Theta_s - \psi/2.
\ee
Here $ \Phi_c $ and $ \Theta_s $ are the charge and the (dual) spin phases, their
gradients are coupled to charge- and spin source fields, respectively \cite{DualFields}:
$ {\cal L}_{h} = h_c \partial_x \Phi_c + h_s \partial_x \Theta_s $.
After shifting the bosonic phases, terms $ \propto h_c \partial_x \alpha / 2 $ and
$ \propto h_s \partial_x \psi / 2 $ arise in the Lagrangian. Finally, we can return to the
fermionic variables:
\be
\label{LagrLE-rot}
   {\cal L}^{\rm (sl)} \simeq {\cal L}_{\rm e} + {\cal L}_{\rm bs}^{\rm (sl)} \! |_{\alpha,\psi=0}
                         + \!\!\!\!\!\! \sum_{2\Phi=\alpha,\psi} \!\!\!\! {\cal L}_{\rm TL}(\Phi,v_F) + {\cal L}_{\rm WZ} .
%
\ee
Here $ {\cal L}_{\rm TL} (\Phi,v) = [(\partial_\tau \Phi)^2 + (v \, \partial_x \Phi)^2]/\pi v $
is the Lagrangian of the Tomonaga-Luttinger Liquid (TLL), i.e., the anomaly \cite{Cross}.

For fixed values of $ \{ \theta, \alpha_{\parallel} \} $, the fermionic  spectrum consists  of the
four Dirac  modes with the masses given by:
\be
\label{GapsPM}
   m_{\pm}^2 = (\tilde{s} \rho_s / 2 )^2
                         \left( \sqrt{J^2_\perp \cos^2\theta + J^2_{z}\sin^2\theta} \pm J_\perp \right)^2.
\ee
Integrating out the gapped fermions, we get the contribution to the ground state energy:
\be
\label{GSen-EA}
  E_{\rm GS} = - \frac{\xi_0}{2 \pi v_F} \sum_{\chi=\pm} m_\chi^2 \ln[t/|m_\chi| ] + o(J_\perp^2,J_z^2).
\ee
If $ J_z > J_\perp $, $ E_{\rm GS} $ has minima at $ \theta = \pi/2, \alpha_\parallel = 0 $;
small fluctuations around the minima read:
\be
  \delta E_{\rm ea} /  {\cal E} \approx
                   (J_{z}^2-J_\perp^2)  \cos^2(\theta) + (J_{z}^2+J_\perp^2) \sin^2(\alpha_\parallel) ,
\label{E}
\ee
where $ {\cal E} \equiv \ln\left( t / J \right) (s \rho_s)^2 \xi_0 / 4 \pi v_F $ and
we do not distinguish between $J_{z}$ and $J_\perp$ in the logarithm.
Using Eqs.(\ref{WZ},\ref{E}) and integrating over the Gaussian fluctuations
of the angles, we can find  parameters of $ {\cal L}_{\rm TL}(\alpha) $
which are renormalized due to the coupling of the spin wave to the gapped
fermions: $ \, 4 v_\alpha / v_F = K_\alpha \ll 1 $ \cite{LuttParam}.
The LE Lagrangian for the EA anisotropy is \cite{Sublead}:
\be
\label{LagrEA}
   {\cal L}_{\rm ea} = {\cal L}_{\rm TL}(\psi,v_F)/4 + {\cal L}_{\rm TL}( \alpha,v_\alpha )/K_\alpha + {\cal L}_h^{\rm (ea)} .
\ee
$ {\cal L}_{\rm ea} $ corresponds to two U(1)-symmetric TLL models with the slow charge, $\alpha$,
and the fast spin, $ \psi $, bosonic modes.

{\it Breaking Z$_2$ symmetry}:
If $ J_z \gg J_\perp $, then $ m_+ \simeq m_- $, all fermionic modes have (almost) the same gap $\sim J_z$, cf.
Eq.(\ref{CouplLagrSlowParam}).
Mass $ m_- $ progressively shrinks towards the SU(2) symmetric point of the quantum
phase transition where $ \, m_- = 0 $ and one subsystem of the helical fermions becomes gapless.
Our approach looses its validity at $ m_- \to 0 $. We  leave a description of the SU(2)
symmetric point  for future studies and consider instead the case of the strong EP
anisotropy $ J_z \ll J_{\perp}$.

{\it EP anisotropy, $ J_z \ll J_{\perp}$}:
To make the  consideration transparent, we put $ J_z \to 0 $ and rewrite
Eq.(\ref{CouplLagrSlowParam}) as a sum of helical contributions:
\bea
\label{Helic-1}
   {\cal L}_{\rm bs}^{\rm (H1)} & = & \tilde{s} \rho_s
            J_\perp R_{\uparrow}^\dagger \cos^2\left( \theta/2 \right) \re^{-\ri (\psi + \alpha)} L_{\downarrow} + H.c. \\
\label{Helic-2}
   {\cal L}_{\rm bs}^{\rm (H2)} & = & - \tilde{s} \rho_s
            J_\perp R_{\downarrow}^\dagger \sin^2\left( \theta/2 \right) \re^{\ri (\psi - \alpha)} L_{\uparrow} + H.c.
\eea

If $ \, \theta \simeq \pi/2 $, both helical sectors have a gap though
the coupling constant $ J_\perp $ is effectively decreased because $ \sin^2\left( \theta/2 \right) \simeq
\cos^2\left( \theta/2 \right) \simeq 1/2 $. If $ \, \theta \simeq 0, \pi $, only one helical sector acquires
the gap $ m = m_+(J_z=0,\alpha_\parallel=\theta=0) $,
and $ \, J_\perp \, $ is not suppressed because either $ \sin^2\left( \theta/2 \right) \simeq 1 $ or
$ \cos^2\left( \theta/2 \right) \simeq 1 $. Since the contribution of the gapped fermions
to the ground state energy is negative and quadratic in the gap, Eq.(\ref{GSen-EA}), we conclude
that $ \, \theta = \pi/2 $ yields maximum of the energy and two (degenerate) minima
are $ \, \theta = 0, \pi $. Thus, the Z$_2$ symmetry between the helical subsystems is
spontaneously broken confirming that the SU(2) symmetric point is the point of a quantum
phase transition \cite{DiscSym}.

Let us consider the configuration $ \theta \simeq 0 $ where
only $ {\cal L}_{\rm bs}^{\rm (H1)} $ yields
the femionic gap \cite{SecondMin}. One can straightforwardly estimate that contributions of the gapped
and the gapless fermions to fluctuations of the ground state energy are of order $ \sim (J_\perp^2/v_F)
\sin^2(\theta/2) $ and $ \sim (J_\perp^2/v_F) \sin^4(\theta/2) $, respectively. The latter is
subleading, it is beyond our accuracy and must be neglected. Thus,  $ {\cal L}_{\rm bs}^{\rm (H2)} $
is irrelevant for the effective LE theory and must be neglected too. The combination
$ \psi - \alpha $ becomes redundant and $ \psi $ in the combination $ \psi + \alpha $ [see
Eqs.(\ref{WZ},\ref{Helic-1})] can be absorbed in $ \alpha $: $ \psi + \alpha \to \alpha $ \cite{ParamNo2}.
Now, we can proceed very similar to the case of the EA anisotropy: (a)
eliminate the shifted spin phase $ \alpha  $ from $ {\cal L}_{\rm bs}^{\rm (H1)} $ with the help of
the transformation
\be
\label{Shift-EP}
   \tilde{\Phi}_c =  \Phi_c + \alpha/2, \
   \tilde{\Theta}_s =  \Theta_s - \alpha/2 ;
\ee
(b) integrate out massive helical fermions and obtain the fermionic energy close to its minima:
\be
  \delta E_{\rm ep} / {\cal E} \simeq J_\perp^2 [ \sin^2(\theta/2) + \sin^2(\alpha_\parallel) / 2 ] \, ;
\nonumber
\ee
(c) integrate out small quadratic fluctuations of angles around the stationary value;
and (d) bosonize fermions from the gapless helical sector by using the Abelian phase
$ \Phi_{\rm H} $. These steps yield the effective Lagrangian for the case of the EP \cite{Sublead}:
\bea
\label{LagrEP}
   {\cal L}_{\rm ep} & = &  {\cal L}_{\rm TL}( \Phi_{\rm H}, v_F )/2 + {\cal L}_{\rm TL}( \alpha,v'_\alpha )/K'_\alpha
                            + {\cal L}_h^{\rm (ep)} ;
\eea
where $ 4 v'_\alpha / v_F =  K'_\alpha \ll 1 $ \cite{LuttParam}.
Similar to the EA anisotropy, $ {\cal L}_{\rm ep} $ corresponds to two U(1)-symmetric TLL
models with the fast, $ \Phi_{\rm H} $, and the slow, $ \alpha $, bosonic modes. However, as
we discuss below, the effective theories with- and without the helical symmetry have
different transport properties if a disorder is added.

To conclude this section, we note that Eqs.(\ref{Helic-1},\ref{LagrEP}) are equivalent to their counterparts
describing a helical edge mode in the TI with an array of the Kondo impurities \cite{AAY,Yevt-Helical}.
In our case, however, this helical mode has emerged as a result of spontaneous symmetry breaking.

{\it Density correlation functions and effects of the disorder}: Let us keep source terms for the charge
sector: $  {\cal L}_{h}^{\rm (ea)} = h_c \, \p_x \alpha / 2 ; \ {\cal L}_{h}^{\rm (ep)} = h_c ( \p_x \Phi_{\rm H}
+  \p_x \alpha / 2) $. The charge density-density correlation function at given frequency and momentum
reads as:
\bea
  {\cal C}_{\rm ea} \propto Q^2 \langle \alpha^* \alpha\rangle ; \
  {\cal C}_{\rm ep} \propto Q^2
                                    \Bigl( \langle \Phi_{\rm H}^* \Phi_{\rm H} \rangle  +
                                    \langle \alpha^* \alpha \rangle / 4 \Bigr) .
\eea
$  {\cal C}_{\rm ea,ep} $ with  Lagrangians $ {\cal L}_{\rm ea, ep} $
correspond to the ideal metallic transport. In the EA case, it is supported
by the slow CDW with the small compressibility, $ K_\alpha $. $ \,  {\cal C}_{\rm ep} \, $ contains the
contribution from the helical quasiparticles with the bare velocity and from the slow collective wave with
the small compressibility, $ K'_\alpha $ \cite{Drude}.

The coupling of  backscattering spinless impurities to the fermions is
described by:
\be
\label{PotDis}
  V_{\rm dis}[g] = g(x) \Psi^\dagger(I\otimes\tau^\dagger)\Psi + H.c.
\ee
Here $g(x)$ is the smooth $2k_F$-component of the scalar random
potential. We use the model of the Gaussian white noise: $ \langle
g^{1,2} \rangle_{\rm dis} = 0; \langle g(x_1) g^*(x_2) \rangle_{\rm dis}
= {\cal D} \delta(x_1 - x_2) $, assuming that the disorder is weak,
$ {\cal D} \ll (m_{\pm}, m ) v_F $, and it cannot change the gaps.
After shifts Eq.(\ref{Shift},\ref{Shift-EP}), the potential $ g $ acquires the
phase factor:  $ g \to g \times \re^{\ri\alpha/2} $. Thus,
the backscatterering impurities are coupled to all gapless charge carriers
(collective waves and helical fermions).

To figure out whether the disorder may lead to localization, we perform
the disorder averaging and integrate out the massive fermions \cite{CollectiveLoc}.
The relevant terms appear only in $ {\cal D}^2$-order and  have a different
form in EA and EP phases. In the first case,  $ {\cal D}^2$ couples directly
to $ \exp(\ri\alpha) $; in the EP phase, it couples to $ R^+_\sigma L_{-\sigma}
\exp(\ri\alpha) $. The latter fact is related to impossibility of single particle
backscattering in the phase with broken helicity. The power counting
indicates the parametric difference in the localization radius in both phases:
$ L^{\rm (loc)}_{\rm ea} / L^{\rm (loc)}_{\rm ep} \sim K_\alpha ({\cal D} / 
v_F m)^{4/3} \ll 1 , $ with $ 
L^{\rm (loc)}_{\rm ep} \sim (v_F/m) \left( v_F m / {\cal D}
\right)^{2} $.

Localization can block the dc transport if a sample size $ L $ is large:
$ L \gg L^{\rm (loc)} $. We thus conclude that the ballistic transport in
the phase with broken helical symmetry acquires the symmetry
protection up to the parametrically large scale $ L^{\rm (loc)}_{\rm ep} $.
This conclusion holds true as long as the U(1) symmetry in the spin sector
is respected. Breaking the U(1) spin symmetry (e.g. after introducing an
anisotropy in the XY-plane) allows the direct backscattering of all fermions
and removes protection of the ideal transport in the EP phase
[cf. localization of the helical edge modes of the TIs \cite{AAY} in the
absence of the U(1) spin symmetry].

{\it Finite temperature effects in the clean case}:
All previous calculations have been done for zero temperature, $ T = 0 $. They can be generalized for
$T\neq 0$ provided $ T $ is smaller than the fermionic gaps. Finite temperature  restores a broken helical
symmetry at $ J_z < J_\perp $ since thermal fluctuations produce  domains
with opposite helicity. When the spin configuration interpolates  between the  phases with  different
helicity there is an energy increase  of the order of the difference between the energy in the unstable
state (with $ \theta \simeq \pi/2 $) and the energy of one of the ground states (with $ \theta \simeq 0 $
or $ \pi $).
Thus, we can estimate the energy of the domain wall as $ E_{\rm wall} \sim m^2 \xi_0 / v_F  $,
cf.Eq.(\ref{GSen-EA}).
The maximal number of the domain walls in the system of the size $ L $ reads as
$ L m / v_F $. If $ T \ll  E_{\rm wall} $, it becomes exponentially suppressed: $ N_{\rm wall}
\sim  L m / v_F \exp(-E_{\rm wall}/T) $. If  $ N_{\rm wall} > 1 $, the walls appear and block the
quasiparticle transport  since the electrons with a given  helicity are massless only in one domain
and massive in the other (neighboring) one. Hence the electrons are reflected from domain
boundaries. On the other hand, an influence of the domain wall on the field $\alpha$ is reduced
to a jump in the Luttinger parameter $ K'_\alpha $ which cannot affect the dc conductance,
cf. Ref.\cite{affleck}. Thus, we arrive at the conclusion that the dc transport in the phase
with the broken helical symmetry will remain ballistic even at finite temperatures.
Temperature effects in the disordered case deserve a separate study because of a complicated
interplay between formation of the domain walls and many-body (de)localization of collective
waves \cite{MBL}.

{\it Validity}:
The effective LE theory, Eqs.(\ref{LagrEA},\ref{LagrEP}), is valid at energies below the
smallest  fermionic gap, $ m_- $ and $ m $ for the EA and the EP anisotropy, respectively.
Since $ m_- $ vanishes at the SU(2) symmetric point, the approach fails in the vicinity of the quantum
critical point. Quickly oscillating contributions $ \propto e^{\pm 2 i k_F x} $, which we
neglected, are generically unable to change the physics at the large distances:  If the Kondo chain
is close to incommensurability the quickly-oscillating exponentials can be treated as random variables,
cf. Ref.\cite{KL-Honner-Gulacsi}. We note, however, that, in the most interesting case of the broken
helical symmetry, the amplitude of the oscillating terms is suppressed in the vicinity of the classical
spin configuration, $ \theta \simeq 0 $, as $ \sim (\xi_0 J_\perp^2/v_F) \sin^4(\theta/2) $ [see the discussion
of the derivation of Eq.(\ref{LagrEP})] which is squared after averaging over the random fluctuations,
i.e.,  becomes negligible.


{\it Conclusions}:
We have demonstrated that the dc charge transport in the Kondo chain model (\ref{KCh})
with the U(1) symmetry of spins remains ballistic in long samples, $ L <
L^{\rm (loc)}_{\rm ep} $ even in the presence of the potential
disorder when the anisotropy of the exchange interaction is of the easy plane type.
Due to the spontaneous breaking of the Z$_2$ symmetry the current is carried by quasiparticles
possessing a particular helicity (i.e. whose spin and chirality are locked) and by composite
spin-fermion  collective modes. In the presence of the U(1) spin symmetry, all gapless modes
are protected from simple backscattering by the mechanism similar to that in noninteracting TIs. We
emphasize that the symmetry protected transport in our model results from interaction
many-body effects instead of the coupling to the non-interacting and topologically non-trivial bulk.
In the case of the easy axis anisotropy, the helical symmetry is respected.
The quasiparticles are fully gapped and the transport is carried solely by the
collective modes. The slow CDWs do not posses the symmetry protection:
the potential disorder can pin them and render the Kondo chain insulating.

\begin{acknowledgments}
A.M.T. acknowledges the hospitality of Ludwig Maximilians University where this work was done.
A.M.T. was supported by the U.S. Department of Energy (DOE), Division of Materials Science, under
Contract No. DE-AC02-98CH10886. O.M.Ye. acknowledges support from the DFG through SFB
TR-12, and the Cluster of Excellence, Nanosystems Initiative Munich. We are grateful to Vladimir
Yudson, Igor Yurkevich  for useful discussions and to Dennis Schimmel for carefully reading the
paper and for his participation in the derivation of the Wess-Zumino term.
\end{acknowledgments}

\bibliography{Bibliography}

\begin{thebibliography}{50}
\bibitem [{NoK()}]{NoKondo}%
  \BibitemOpen
  \href@noop {} {}\bibinfo {note} {This statement means that the spin density
  is sufficiently high so that the Kondo effect is cut by the gap generated by
  backscattering terms in $ \, \hat{H} $, cf. Refs.\cite{AAY,Yevt-Helical}. It
  holds true when the gap exceeds the Kondo temperature.}\BibitemShut {Stop}%
\bibitem [{Hel()}]{HelicConfig}%
  \BibitemOpen
  \href@noop {} {}\bibinfo {note} {This approach is supported by numerical
  studies of the Kondo chain of classical spins which indicate that the phase
  diagram of the system reflects helical spin configurations
  \cite{ClassKondoChain}. The physics, which is governed by the helical effects
  in spins, is also studied in
  Refs.\cite{Loss-Helic1,Loss-Helic2,Been-Helic}.}\BibitemShut {Stop}%
\bibitem [{Sca()}]{ScaleSep}%
  \BibitemOpen
  \href@noop {} {}\bibinfo {note} {We note that the procedure of separation of
  slow and fast variables is standard and is routinely used in RG calculations:
  see, for example, Chapter~1 in Ref.\cite{ATsBook}.}\BibitemShut {Stop}%
\bibitem [{Dua()}]{DualFields}%
  \BibitemOpen
  \href@noop {} {}\bibinfo {note} {Two remaining fields can be integrated
  out.}\BibitemShut {Stop}%
\bibitem [{Cro()}]{Cross}%
  \BibitemOpen
  \href@noop {} {}\bibinfo {note} {While deriving Eq.(\ref{LagrLE-rot}), we
  have neglected the gradient coupling of the fermions and the spin phases, $
  \Psi^\dagger [\partial_{x,\tau}(\alpha,\psi)] \Psi $, which is justified for
  energies below the fermion gaps}\BibitemShut {NoStop}%
\bibitem [{Sub()}]{Sublead}%
  \BibitemOpen
  \href@noop {} {}\bibinfo {note} {While deriving Eqs.(\ref{LagrEA}) and
  (\ref{LagrEP}), we have neglected in $ {\cal L}_{\rm WZ} $ the subleading
  terms $ \propto \sin(\alpha_\parallel) \cos(\theta) \partial_\tau \psi $ and
  $ \propto \sin(\alpha_\parallel) \sin^2(\theta) \partial_\tau \psi $,
  respectively. This is justified for energies below the fermion
  gaps.}\BibitemShut {Stop}%
\bibitem [{Dis()}]{DiscSym}%
  \BibitemOpen
  \href@noop {} {}\bibinfo {note} {We note that helicity is given by the
  product of the spin projection and the electron velocity, $ H = {\rm
  sign}(\sigma v) $. Hence, helical symmetry is discrete and it can be
  spontaneously broken in one dimension at $T=0$. The corresponding order
  parameter is related to the average $ {\cal C}_a = \langle S_b(x)
  S_c(x+\xi_0) \epsilon_{abc} \rangle $: in the phase with the broken symmetry
  $ {\cal C}_z \propto \sin(2 k_F \xi_0) $ and $ {\cal C}_z = 0 $
  otherwise.}\BibitemShut {Stop}%
\bibitem [{Sec()}]{SecondMin}%
  \BibitemOpen
  \href@noop {} {}\bibinfo {note} {The second minima $ \theta \simeq \pi $ can
  be considered similarly and it yields the same effective Lagrangian $ {\cal
  L}_{\rm ep} $, Eq.(\ref{LagrEP}).}\BibitemShut {Stop}%
\bibitem [{Dru()}]{Drude}%
  \BibitemOpen
  \href@noop {} {}\bibinfo {note} {The small compressibility of the bosonic
  modes leads to suppression of the Drude weight which reflects the coupling of
  the spin waves to the gapped fermions \cite{AAY}}\BibitemShut {NoStop}%
\bibitem [{Col()}]{CollectiveLoc}%
  \BibitemOpen
  \href@noop {} {}\bibinfo {note} {All essential details of this calculus are
  presented in Suppl.Mat. No.4.}\BibitemShut {Stop}%
\end{thebibliography}


\protect\pagebreak

\onecolumngrid

\section*{Supplemental Materials}

\subsection{1. Derivation of the Wess-Zumino term}

Here we discuss the subject of spin action which is usually formulated
in the Wess-Zumino form \cite{ATsBook}. This form is invariant under rotations,
however, it requires an integration over an auxiliary variable which is not convenient for our
purposes. We have to find another formulation which would not include the additional
integration and, nevertheless, would allow us to change the basis.

Let us start with the spin being defined as ${\bf S} = s \, {\bf e}_3$, see the paragraph before
Eq.(\ref{NewSpin}) in the main text. Here $ {\bf e}_3$ is one of vectors from the orthonormal
basis $ {\bf e}_{1,2,3} $, an example of the vectors $ {\bf e}_{1,2} $ is given in Eqs.(\ref{Evec-1},\ref{Evec-2}).
The Wess-Zumino term for this representation is well-known:
\be  \label{WZ-1}
   {\cal L}_{\rm WZ}[\theta,\psi] =  \frac{\ri s \rho_s}{\xi_0} \cos(\theta) \p_{\tau} \psi \,  .
\ee
The boundary contribution has been neglected in Eq.(\ref{WZ-1}) and in all equations below since
we are interested in smooth (semiclassical) spin modes. We note that  Eq.(\ref{WZ-1}) is invariant
with respect to all O(2) rotations of the vectors $ \, {\bf e}_{1,2} \perp {\bf e}_3 $:
\be
  \label{GaugeRot}
  {\bf e}_1 = \cos(\beta) \tilde{{\bf e}}_1 +\sin(\beta) \tilde{{\bf e}}_2, \quad
  {\bf e}_1 = -\sin(\beta) \tilde{{\bf e}}_1 +\cos(\beta) \tilde{{\bf e}}_2 .
\ee
This is because $ \beta $ is the gauge angle and it either drops out from Eq.(\ref{WZ-1}) (if
$ \beta = {\rm const}$) or yields an unimportant boundary contribution (if $ \beta $ depends on time).
This allows us to rewrite Eq.(\ref{WZ-1}) by using the vectors  $ \, {\bf e}_{1,2} \, $ from any orthonormal
basis:
\be \label{WZ-1-inv}
   {\cal L}_{\rm WZ}[\theta,\psi] = \frac{\ri s \rho_s}{2 \xi_0}[({\bf e}_2, \p_{\tau}{\bf e}_1) - ({\bf e}_1, \p_{\tau}{\bf e}_2)] =
                                                       \frac{\ri s \rho_s}{\xi_0} ({\bf e}_2, \p_{\tau}{\bf e}_1) .
\ee
The first equality in Eq.(\ref{WZ-1-inv}) can be verified by direct inspection after inserting expressions
(\ref{Evec-1},\ref{Evec-2}) into Eq.(\ref{WZ-1-inv}) and the second one follows from the orthogonality
condition $  ({\bf e}_1, {\bf e}_2) = 0 $. We note that Eq.(\ref{WZ-1-inv}) contains only scalar products
of two vectors and, therefore, it is invariant under the global rotation of the $\{x,y,z\}$-basis:
\be
  \label{XYZ-rot}
  ({\bf e}_j, \p_{\tau}{\bf e}_k) = ({\bf E}_j, \p_{\tau}{\bf E}_k) \quad \mbox{ if }  \quad
  {\bf E}_{1,2} = \hat{R}_{\rm xyz} \, {\bf e}_{1,2}, \quad  \hat{R}_{\rm xyz} \hat{R}^{\rm T}_{\rm xyz} = 1 .
\ee
To avoid confusions, subscriptips of orthogonal matrices show the basis where they operate.

Now we change to the new spin, Eq.(\ref{NewSpin}), and define the new orthonormal basis  $  {\bf e}_{1,23}' $ with:
\be
 \label{E3-new}
     {\bf e}_3' \equiv \roarrow{\cal S} / s =
       \cos({\alpha_\parallel}) [\cos(\alpha_\perp) {\bf e}_1 + \sin(\alpha_\perp){\bf e}_2]  +
                \sin({\alpha_\parallel}) {\bf e}_3 \, .
\ee
Two remaining vectors from the new basis can be chosen, for example, as follows:
\bea
 \label{E1-new}
 {\bf e}_1'  & = & -\sin({\alpha_\parallel})
                             [\cos( \alpha_\perp ) {\bf e}_1 + \sin( \alpha_\perp ){\bf e}_2] +  \cos({\alpha_\parallel}) {\bf e}_3, \\
 \label{E2-new}
 {\bf e}_2' & = & \sin( \alpha_\perp ) {\bf e}_1 - \cos( \alpha_\perp ){\bf e}_2.
\eea
Let us first assume that $ \alpha_{\perp,\parallel} $ does not depend on time. In this simple case, the transformation
(\ref{E3-new}--\ref{E2-new}) is global in the $ {\bf e}_{1,2,3} $ basis but it is local in the $\{x,y,z\}$-basis. The latter
statement results from the rotation of the basis vectors $ {\bf e}_{1,2,3} $. Thus, before using Eq.(\ref{E3-new}--\ref{E2-new}),
we have to rewrite Eq.(\ref{WZ-1-inv}) in the form which is invariant under all possible global rotations. Such a
form reads as
\be
  \label{WZ-2-inv}
   {\cal L}_{\rm WZ}[\theta,\psi] = - \frac{\ri \rho_s}{2 \xi_0} ({\bf S}, {\bf e}_i) (  {\bf e}_j,  \partial_\tau {\bf e}_k ) \epsilon_{ijk},
\ee
where $  \epsilon_{ijk} $ is the antisymmetric tensor. Eq.(\ref{WZ-2-inv}) reduces to Eq.(\ref{WZ-1-inv}) if $ {\bf S} = s \,  {\bf e}_3 $.
It is clearly invariant under the global rotation by the matrix $ \hat{R}_{\rm xyz} $. The invariance under the global rotation by the
matrix  $ \hat{R}_{\rm 123} $ can be easily shown after rewriting the vector $ {\bf S} $ in the $ {\bf e}_{1,2,3} $-basis, substituting
the basis $ ( {\bf e}_{1}', {\bf e}_{2}', {\bf e}_{3}' )^{\rm T} = \hat{R}_{\rm 123} ( {\bf e}_{1}, {\bf e}_{2}, {\bf e}_{3} )^{\rm T} $ into
Eq.(\ref{WZ-2-inv}) and using the identity
\[
   \sum_{i,j,k=1,2,3} [\hat{R}_{\rm 123}]_{ip}  \, [\hat{R}_{\rm 123}]_{jm}  \, [\hat{R}_{\rm 123}]_{kq} \, \epsilon_{ijk} =\epsilon_{pmq} \, .
\]

Following the approach explained in Sect. {\it Continuum limit} of
the main text, we insert
Eqs.(\ref{E3-new}--\ref{E2-new}) into Eq.(\ref{WZ-2-inv}) allowing the angles $ \alpha_{\perp,\parallel} $ to be the
independent variables and substitute $ s \, \sin(\alpha_\parallel) {\bf e}_3 $ for the spin $ {\bf S}$. This procedure
yields Eq.(\ref{WZ}) in the main text. The second substitution takes into account an effective decrease of the slow component
of the spin: after introducing the new rotating frame, the size of the spin will be given by the overlap of this effective
spin with the original one, i.e. by $ s({\bf e}_3, {\bf e}_3') = s \sin( \alpha_{\parallel} ) $, see Fig.\ref{Geom}.


To finalize the discussion of the Wess-Zumino term, we point out a short-cut which allows one to obtain the
answer Eq.(\ref{WZ}) even faster: we can 1) exploit the gauge invariance described in Eq.(\ref{GaugeRot}) and directly
insert the vectors $ {\bf e}_{1,2}'$ into Eq.(\ref{WZ-1-inv}); 2)  allow the angles $ \alpha_{\perp,\parallel} $ to be independent
variables in the actions; 3) do the shift of $ \alpha_\perp $ and omit the oscillatory terms in $ {\cal L}_{\rm WZ} $.


\subsection{2. Alternative derivation of the effective Lagrangian in the case $ J_z = 0 $.}

Let us consider the extreme case of the easy plane anisotropy where $ J_z = 0 $.
Eq.(\ref{CouplLagr}) simplifies to
\be
\label{CouplLagr-Simpl}
   {\cal L}_{\rm bs} = \rho_s J_\perp \re^{2 \ri k_F x} \left( S^+ R^\dagger_\downarrow  L_\uparrow +
                                        S^- R_\uparrow^\dagger  L_\downarrow \right) + H.c.
\ee
here $ S^\pm \equiv S^x \pm \ri S^y $. Now we can use the standard parametrization
of $ \bf S$ by azimuthal and polar angles:
\bea
  {\bf S} & = & s [ \sin(P)\cos(A), \sin(P)\sin(A), \cos(P) ] , \\
  \Rightarrow
  {\cal L}_{\rm bs} & = & s \rho_s J_\perp \cos(P) \re^{2 \ri k_F x} \left( \re^{\ri A} R^\dagger_\downarrow  L_\uparrow +
                                        \re^{- \ri A} R_\uparrow^\dagger  L_\downarrow \right) + H.c.
\eea
(notation for the angles are changed as compared to the main text to avoid confusions)
and note that the slow spin modes can be easily singled out after a shift
\be
  A \to A \pm 2 k_F x .
\ee
The choice of the $\pm$ sign substitutes now the choice of the minima (either $ \theta = 0 $ or $ \theta = \pi $,
see the main text): it breaks the helicity symmetry and corresponds to the strong correlation
of the spins to one or the other sector of the helical fermions. The helical sector, which is not correlated
with the spins, acquires $ 4 k_F $-oscillations and vanishes in the effective Lagrangian for the
backscattering. For example, choosing the plus sign, we obtain
\be
  {\cal L}_{\rm bs}^{\rm (H1)} =  s \rho_s J_\perp \cos(P) \re^{-\ri A} R_{\uparrow}^\dagger L_{\downarrow} + H.c., \quad
  {\cal L}_{\rm bs}^{\rm (H2)} = 0 ;
\ee
with the Wess-Zumino term $ {\cal L}_{\rm WZ} = \ri s \rho_s \xi_0^{-1} \cos(P) \p_\tau A$ and with
the integration measure $ \rcd\{\Omega\} = \sin(P) \, \rcd \{P\} \rcd \{A\} $. This confirms
that the third angle becomes redundant for the effective low-energy theory at $ J_z = 0 $.

\subsection{3. Luttinger parameters $ K_\alpha $ and $ K'_\alpha $}

Calculations described after Eq.(\ref{E}) of the main text yield the following expressions
for the Luttinger parameters:
\be
\label{Compr-a}
  K_\alpha \simeq \xi_0 \frac{\sqrt{J_z^2 + J_\perp^2}}{\pi v_F } \sqrt{\log(t/J)} \ll 1 ; \quad
  K'_\alpha \simeq \frac{4 m \xi_0}{ \pi s \rho_s v_F } \sqrt{\log(t/m)} \ll \! 1 .
\ee

\subsection{4. Derivation of the localization radius, $ L^{\rm (loc)} $.}

In this section, we derive an estimate for the localization radius in the Kondo chain coupled to
spinless backscattering impurities. Firstly, we replicate fields
\be
   V_{\rm dis}[g \, \re^{\ri\alpha/2}] \to \sum_a \sum_\sigma g \re^{\ri \alpha_a/2} R^{\dagger}_{a,\sigma} L_{a,\sigma} , \quad
   \sigma = \uparrow, \downarrow \, ;
\ee
and calculate the Gaussian integral over the random field $ g $. This yields the standard
contribution to the action
\be
  S_{\rm dis} = - \frac{{\cal D}}{2} \int {\rm d}\{ x, \tau_{1,2} \} \sum_{a_{1,2}} \sum_{\sigma_{1,2}}
                           \left( \re^{\ri \alpha/2} (R^{\dagger} L) \right)[{\bf 1}] \times
                           \left( \re^{-\ri \alpha/2} (L^{\dagger} R) \right)[{\bf 2}] + H.c. \, , \quad
                           {\bf n} \equiv \{ x, \tau_n, a_n, \sigma_n \} \, , n = 1,2 \, .
\ee
At the next step, we integrate out massive fermions perturbatively by doing an expansion
in the small parameter $ \, {\cal D} / (m_\pm, m) v_F \ll 1 $. Our goal is to find leading
terms which can result in pinning of all collective charge carriers (EA and EP phases)
and in localization of the massless helical fermions (EP phase). We do it separately
for the two different phases.

\subsubsection{4.1 The EA phase: most relevant terms}

Let us put $ \, J_\perp \to 0 \, $ and introduce a spin-dependent fermionic mass
$ \, m_{\rm ea} (\sigma) = \pm  (\tilde{s} \rho_s / 2) J_z  \, $ where the
plus (minus) correspond to $ \, \sigma = \uparrow \, $ ($ \sigma = \downarrow $).
This allows us to simplify the derivation without loss of generality. The matrix Green's
function for the fermions with a given spin reads:
\be
\label{GFm}
  \hat{G}_m(\sigma) = \left( G^{(0)}_R G^{(0)}_L - m_{\rm ea}(\sigma)^2  \right)^{-1}
      \left(
         \begin{array}{cc}
                \left( G^{(0)}_L \right)^{-1}  & -  m_{\rm ea}(\sigma)          \\
                    -  m_{\rm ea}(\sigma)     &  \left( G^{(0)}_R \right)^{-1}
         \end{array}
      \right) ;
\ee
where $ G^{(0)}_{R,L} $ are the Green's functions of free chiral particles. It is important
that $ \hat{G}_m $ is short ranged and it decays beyond the time scale $ 1/m_{\rm ea} $
(or beyond the coherence length $ \xi_{\rm ea} \equiv v_f/m_{\rm ea} $).

\begin{figure}[h]
   \includegraphics[width=0.5 \textwidth]{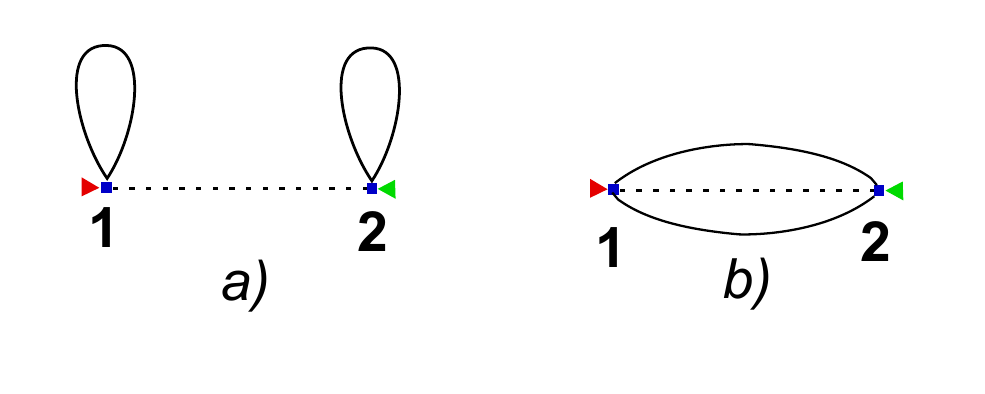}
   \vspace{-0.5 cm}
   \caption{\label{EA-d1}
        First order diagrams $ \, O({\cal D}^1) \, $ for the EA phase. Red (green) triangulars
        denote $ \, \re^{\ri \alpha/2} \, $ ($ \, \re^{-\ri \alpha/2} \, $) with arguments of either the 1st
        or the 2nd vertex; dashed lines are the disorder correlation functions, solid lines stand
        for Green's functions of the massive fermions.
                 }
\end{figure}

Leading terms of the order of $ \, O({\cal D}^1) \, $ are given by $ \langle S_{\rm dis} \rangle $
where brackets mean that the massive fermions are integrated out. The corresponding
diagrams are shown in Fig.\ref{EA-d1}. It is easy to check that the diagrams from Fig.\ref{EA-d1}-a
cancel out after summation over spins indices because $ \, m_{\rm ea} (\uparrow) = - m_{\rm ea}
(\downarrow) $. The diagrams from Fig.\ref{EA-d1}-b are trivial since $ \hat{G}_m $ is diagonal
in the replica space and the spin phase $ \alpha $ is smooth on the scale $ 1/m_{\rm ea} $; therefore,
\be
  \label{Canc-a}
  \re^{\ri \alpha[{\bf 1}]/2} \re^{-\ri \alpha[{\bf 2}]/2} \simeq 1 \, ,
\ee
with some small gradient corrections
which are unable to yield pinning.

\begin{figure}[h]
   \includegraphics[width=0.5 \textwidth]{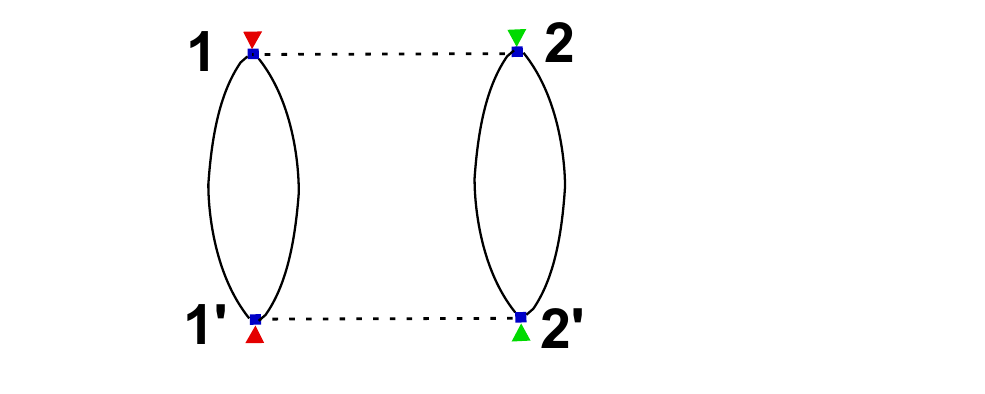}
   \vspace{-0.5 cm}
   \caption{\label{EA-d2}
        A typical non-trivial diagram, $ D^{(2)}_{\rm ea} $, of the order $ \, O({\cal D}^2) \, $
        for the EA phase; notations are explained in the caption of Fig.\ref{EA-d1}.
                 }
\end{figure}

Sub-leading terms of the order of $ \, O({\cal D}^2) \, $ are given by $ \langle S_{\rm dis}
S_{\rm dis} \rangle $ which generates a lot of diagrams. We leave a detailed analysis for
the future and calculate only one typical diagrams which survives after all summations
and is able to generate pinning. An example of such a diagram is shown in Fig.\ref{EA-d2}.

Neglecting unimportant numerical factors, the analytical expression for the diagram
from Fig.\ref{EA-d2} reads as:
\be
\label{D2-EA}
  D^{(2)}_{\rm ea} \propto
  {\cal D}^2 \sum_{a_{1,2}} \int {\rm d} \{ x, x' ; \tau_{1,2}, \tau_{1,2}' \}
  \re^{\ri \left( \alpha[{\bf 1}] -  \alpha[{\bf 2}] \right)}
  \left[ \hat{G}_m({\bf 1,1'}) \right]_{1,2} \left[ \hat{G}_m({\bf 1',1}) \right]_{1,2}
  \left[ \hat{G}_m({\bf 2,2'}) \right]_{1,2} \left[ \hat{G}_m({\bf 2',2}) \right]_{1,2} \, .
\ee
Here, we have taken into account the the diagonal structure of $ \hat{G}_m $ results
in $ a_1 = a_1'; a_2 = a_2' $
and fused together slow spin phases, for instance: $ \alpha[{\bf 1}] + \alpha[{\bf 1'}] \simeq
2 \alpha[{\bf 1}] $. Now we note that $ \hat{G}_m({\bf 1,1'}) = \hat{G}_m({\bf 1 - 1'}) $
and integrate over all primed variables:
\be
\label{EA-SG}
  D^{(2)}_{\rm ea} \propto
  \frac{ \tilde{{\cal D}}_0 }{\xi_{\rm ea}^2}
  \sum_{a_{1,2}}
  \int {\rm d} \{ x; \tau_{1,2} \} \re^{\ri \left( \alpha[{\bf 1}] -  \alpha[{\bf 2}] \right)} \, ; \quad
  \tilde{ {\cal D} }_0 \equiv {\cal D} \left( \frac{{\cal D}}{v_F m_{\rm ea}} \right) .
\ee
The structure of Eq.(\ref{EA-SG}) corresponds to the disordered Sine-Gordon model
which appears in the theory of the usual TLL \cite{Giamarchi}. The effective disorder strength
$  \tilde{ {\cal D} } $ is renormalized and obeys the well-known RG equation \cite{GiaSchulz}:
\be
\label{RG-EA}
  {\mbox EA}: \quad
  \partial_{\log} \log( \tilde{ {\cal D} } ) = 3 - 2 K_\alpha \simeq 3 \, , \quad
  \tilde{ {\cal D} }(\xi_{\rm ea}) = \tilde{{\cal D}}_0 \, ;
\ee
the second equality of Eq.(\ref{RG-EA}) has been obtained by using Eq.(\ref{Compr-a}).

\subsubsection{4.2 The EP phase: most relevant terms}

\begin{figure}[h]
   \includegraphics[width=0.5 \textwidth]{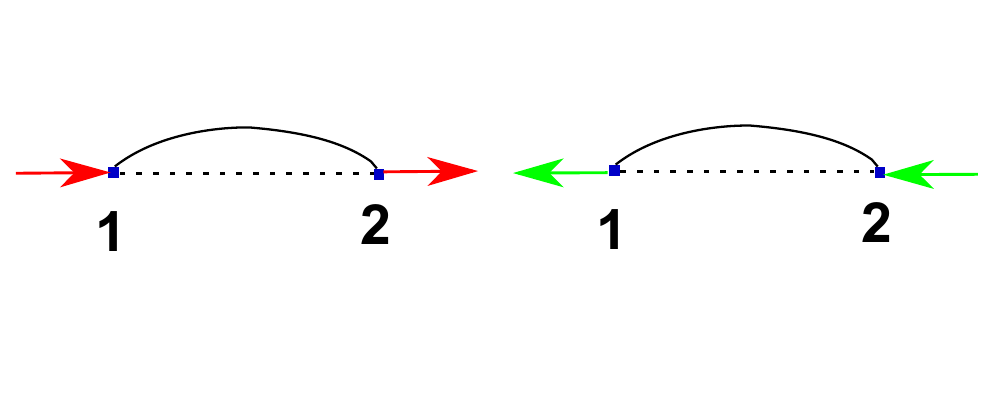}
   \vspace{-1.25 cm}
   \caption{\label{EP-d1}
        Two typical examples of first order diagrams $ \, O({\cal D}^1) \, $ for the EP phase.
        Incoming red (green) arrows
        denote the product of smooth fields $ \, \re^{\ri \alpha/2} L_{\uparrow} \, $ ($ \, \re^{-\ri \alpha/2}
        R_{\downarrow} \, $) with arguments of either the 1st or the 2nd vertex; outgoing arrows
        denote the conjugated product, dashed lines are the disorder correlation functions, solid
        lines stand for Green's functions of the massive helical fermions.
                 }
\end{figure}

We start again from the leading diagrams generated by $ \langle S_{\rm dis} \rangle $.
The principal difference of the EP phase from the EA one is that the matrix Green's
function, Eq.(\ref{GFm}), corresponds now to the massive fermions with a given helicity.
This changes the structure of the first order diagram, see Fig.\ref{EP-d1}. All these
diagrams correspond to forward-scattering of the massless helical fermions and
they contain only small gradients of the phase $ \, \alpha $, cf. Eq.(\ref{Canc-a}) and
its explanation. Thus, the leading diagrams are trivial and they cannot yield localization,
the sub-leading diagrams must be considered.

\begin{figure}[h]
   \includegraphics[width=0.5 \textwidth]{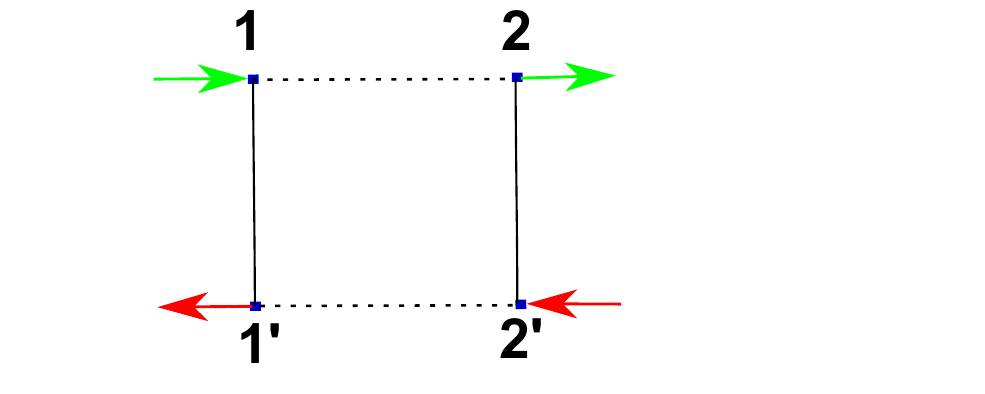}
   \vspace{-0.5 cm}
   \caption{\label{EP-d2}
        A typical non-trivial diagram, $ D^{(2)}_{\rm ep}$, of the order $ \, O({\cal D}^2) \, $ for the EP phase;
        notations are explained in the caption of Fig.\ref{EP-d1}.
                 }
\end{figure}

$ \langle S_{\rm dis} S_{\rm dis} \rangle $ generates 16 diagrams with back-scattering of
the massless fermions and exponentials of the phase $ \alpha $ which do not cancel, see Fig.\ref{EP-d2}
Neglecting unimportant numerical factors, the analytical expression for the diagram
from Fig.\ref{EP-d2} reads as:
\be
  D^{(2)}_{\rm ep} \propto
  {\cal D}^2 \sum_{a_{1,2}} \int {\rm d} \{ x, x' ; \tau_{1,2}, \tau_{1,2}' \}
  \re^{\ri \left( \alpha[{\bf 1}] -  \alpha[{\bf 2}] \right)} \,
  L^\dagger_\downarrow[ {\bf 2} ] R^\dagger_\uparrow[ {\bf 1} ] \, L_\downarrow[ {\bf 1} ] R_\uparrow[ {\bf 2} ] \,
  \left[ \hat{G}_m({\bf 1,1'}) \right]_{1,2}
  \left[ \hat{G}_m({\bf 2,2'}) \right]_{1,2} \, ;
\ee
see explanations after Eq.(\ref{D2-EA}) and note the $ m $ must be substituted for $ m_{\rm ea} (\sigma) $
in $ \hat{G}_m $.  Calculating integrals over all primed variables, we find:
\be
\label{EP-SG}
  D^{(2)}_{\rm ep} \propto
  \bar{{\cal D}}_0
  \sum_{a_{1,2}} \int {\rm d} \{ x ; \tau_{1,2} \}
  \re^{\ri \left( \alpha[{\bf 1}] -  \alpha[{\bf 2}] \right)} \,
  L^\dagger_\downarrow[ {\bf 2} ] R^\dagger_\uparrow[ {\bf 1} ] \, L_\downarrow[ {\bf 1} ] R_\uparrow[ {\bf 2} ] \, , \quad
  \bar{ {\cal D} }_0 \equiv {\cal D} \left( \frac{{\cal D}}{v_F m} \right) .
\ee
This equation also can be reduced to the form of Eq.(\ref{EA-SG}) if remaining fermions are bosonized
and we explicitly single out new charge- and spin- density waves. However, the RG equation for
$ \bar{ {\cal D} } $ can be obtained without such a complicated procedure with the help of the
power counting. Firstly we note that the scaling dimension of each back-scattering term in Eq.(\ref{EP-SG}),
$ L^\dagger R $ and $ R^\dagger L $, is 1. The anomalous dimension of each exponential, $ \re^{\pm \ri
\alpha} $, is $ K_\alpha' \ll 1 $. The normal dimension in Eq.(\ref{EP-SG}) is 3 which comes from three-fold
integral. Combining these dimensions together and neglecting small $ K_\alpha'  $, we find
\be
\label{RG-EP}
  {\mbox EP}: \quad
  \partial_{\log} \log( \bar{ {\cal D} } ) = 3 - 2 \times 1 + O(K_\alpha) \simeq 1 \, ; \quad
  \bar{ {\cal D} }(\xi_{\rm ep}) = \bar{{\cal D}}_0 \, , \ \xi_{\rm ep} = v_F / m \, .
\ee

\subsubsection{4.3 Comparison of the localization radius in different phases}

The solution of the RG equations, Eqs.(\ref{RG-EA},\ref{RG-EP}), reads as
\be
  \tilde{\cal D}(x) = \tilde{\cal D}_0 \left( \frac{x}{\xi_{\rm ea}} \right)^3, \quad
  \bar{\cal D}(x) = \bar{\cal D}_0 \frac{x}{\xi_{\rm ep}} \, ;
\ee
with $ \xi_{\rm ep} = v_F / m $. The localization radius is defined as a scale on
which the renormalized disorder becomes of the order of the cut-off:
\be
\label{LlocEq}
   \tilde{\cal D}\left(L^{\rm (loc)}_{\rm ea}\right) =
      K_\alpha v_\alpha^2 / \xi_{\rm ea} \sim K_\alpha^3 v_F^2 / \xi_{\rm ea} \, ; \quad
   \bar{\cal D}\left(L^{\rm (loc)}_{\rm ep}\right)  = v_F^2 / \xi_{\rm ep} \, .
\ee
The additional small factor $ K_\alpha $ is the equation for $ L^{\rm (loc)}_{\rm ea} $ can be justified
with the help of the standard optimization procedure \cite{Giamarchi} where $ L^{\rm (loc)} \, $ is
defined as a spatial scale on which the typical potential energy of the disorder becomes equal to
the energy governed by the term $ \, \propto (\partial_x \alpha)^2 \, $ in the Lagrangian
$ {\cal L}_{\rm ea} $, Eq.(\ref{LagrEA}).

Definitions Eq.(\ref{LlocEq}) result in
\be
  L^{\rm (loc)}_{\rm ea} \sim  \xi_{\rm ea} K_\alpha \left(  \frac{v_F^2}{ \xi_{\rm ea} \tilde{\cal D}_0 } \right)^{1/3} \!\!\! \sim
                                               \xi_{\rm ea} K_\alpha \left( \frac{v_F m_{\rm ea}}{ {\cal D} } \right)^{2/3} \!\!\! ; \quad
  L^{\rm (loc)}_{\rm ep} \sim \frac{v_F^2}{\bar{\cal {D}}_0} \sim
                                              \xi_{\rm ep} \left( \frac{v_F m}{{\cal D}} \right)^2 \, .
\ee
Assuming $ \xi_{\rm ea} \sim \xi_{\rm ep}$ and $ m_{\rm ea} \sim m $, we obtain
\be
  \frac{L^{\rm (loc)}_{\rm ea}}{L^{\rm (loc)}_{\rm ep}} \sim K_\alpha \left( \frac{{\cal D}}{v_F m} \right)^{4/3} \ll 1 \, .
\ee
This demonstrates that the strong suppression of localization can occur in the EP phase where the
helical symmetry is broken.

We note in passing that the scaling exponent of $ \bar{\cal D}(x) $ is the same as in
the case of non-interacting 1d fermions but suppression of localization in the EP
phase is reflected by the additional large factor $  v_F m / {\cal D} $ in the expression
for the localization radius $ L^{\rm (loc)}_{\rm ep} $.

\end{document}